\input{aipcheck}
\documentclass[final]{aipproc}
\layoutstyle{6x9}
\usepackage{amssymb}

\newcommand{\ftn}{\footnotesize}

\def\openep{\leavevmode\hbox{\normalsize$\iota$\kern-3.8pt$^$-}}
\def\vtau{\leavevmode\hbox{\normalsize$\tau$\kern-5.pt$\iota$}}
\def\vtauf{\leavevmode\hbox{\ftn$\tau$\kern-4.pt$\iota$}}

\def\beq{\begin{equation}}
\def\eeq{\end{equation}}
\def\bea{\begin{eqnarray}}
\def\eea{\end{eqnarray}}

\begin{document}

\title{F-Term Hybrid Inflation Followed by Modular Inflation}


\author{C. Pallis}{
address={Departamento de Fisica Aplicada, Universidad de Huelva,
21071 Huelva, SPAIN\\ Department of Physics, University of Patras,
Panepistimioupolis, GR-26500 Patras, GREECE } }

\begin{abstract}
We consider the well motivated model of the (standard)
\emph{supersymmetric} (SUSY) \emph{F-term hybrid inflation} (FHI)
which can be realized close to the \emph{grand unification} (GUT)
scale. The predicted scalar spectral index $n_{\rm s}$ cannot be
smaller than $0.98$ and can exceed unity including corrections
from minimal \emph{supergravity} (SUGRA), if the number of
e-foldings corresponding to the pivot scale $k_*=0.002/{\rm Mpc}$
is around 50. These results are marginally consistent with the
fitting of the \emph{five-year Wilkinson microwave anisotropy
probe} (WMAP5) data by the standard power-law cosmological model
with cold dark matter and a cosmological constant, $\Lambda$CDM.
However, $n_{\rm s}$ can be reduced by restricting the number of
e-foldings that $k_*$ suffered during FHI. The additional
e-foldings required for solving the horizon and flatness problems
can be generated by a subsequent stage of fast-roll [slow-roll]
\emph{modular inflation} (MI) realized by a string modulus which
does [does not] acquire effective mass ($m_s|_{\rm eff}$) before
the onset of MI.

\end{abstract}

\maketitle

\section{Introduction}\label{intro}

We focus on the model of the (standard) SUSY FHI~\cite{hybrid}
which can be realized \cite{hsusy} adopting the superpotential
$W=\kappa S\left(\bar \Phi\Phi-M^2\right)$ where $\Phi$,
$\bar{\Phi}$ is a pair of left handed superfields belonging to
non-trivial conjugate representations with dimensionality ${\sf
N}$ of a gauge group $G$, and reducing its rank by their
\emph{vacuum expectation values} (VEVs), $S$ is a $G$ singlet left
handed superfield and the parameters $\kappa$ and $M$ can be made
positive.

$W$ leads to the spontaneous breaking of $G$ since from the
emerging scalar potential $V_{\rm F}= \kappa^2M^4\left(({\sf
\Phi}^2-1)^2+2{\sf S}^2{\sf \Phi}^2\right)$ where ${\sf
\Phi}=|\Phi|/M$ and ${\sf S}=|S|/M$ (we use the same symbol for
the superfields and their scalar components) we can deduce that
the vanishing of the F-terms gives the VEVs of the fields in the
SUSY vacuum, $\langle {\sf S}\rangle=0$ and $\langle{\sf
\Phi}\rangle=1$ (the vanishing of the D-terms implies that
$\vert\bar{\Phi}\vert=\vert\Phi\vert$). $W$ gives also rise to
FHI. This is due to the fact that, for ${\sf S}>1$, the direction
with ${\sf \Phi}=0$ is a valley of local minima with constant
$V_{\rm F}$. The general form of the potential which can drive FHI
reads
\beq\label{Vol} V_{\rm HI}\simeq\kappa^2 M^4+{\kappa^4 M^4 {\sf
N}\over32\pi^2}\left(2 \ln {\kappa^2\sigma^2 \over
2Q^2}+3\right)+\kappa^2 M^4{\sigma^4\over8m^4_{\rm
P}},~~\mbox{with}~~\sigma=\sqrt{2}S.\eeq
Here, the 1rst term is the dominant contribution to $V_{\rm HI}$,
the 2nd term (with $Q$ being an arbitrary renormalization scale)
is the contribution to $V_{\rm HI}$ due to logarithmic radiative
corrections originating from the SUSY breaking on the inflationary
valley and the 3rd term (with $m_{\rm P}\simeq 2.44\times
10^{18}~{\rm GeV}$) is the SUGRA correction \cite{senoguz} to
$V_{\rm HI}$, assuming minimal K\"ahler potential.

Under the assumption that the cosmological scales leave the
horizon during FHI and are not reprocessed, we can extract:
\begin{itemize}

\item The number of $e$-foldings $N_{\rm HI*}=\:\frac{1}{m^2_{\rm P}}\;
\int_{\sigma_{\rm f}}^{\sigma_{*}}\, d\sigma\: \frac{V_{\rm
HI}}{V'_{\rm HI}}$ that $k_*$ suffered during FHI, where prime
means derivation with respect to $\sigma$, $\sigma_{*}$ is the
value of $\sigma$ when the scale $k_*$ crossed outside the horizon
of FHI and $\sigma_{\rm f}$ is the value of $\sigma$ at the end of
FHI, which coincides practically with the end of the phase
transition $\sigma_{\rm c}=M/\sqrt{2}$.

\item The power spectrum of the
curvature perturbations at $k_{*}$, $P_{\cal
R*}={1\over2\sqrt{3}\, \pi m^3_{\rm P}}\; \left.{V_{\rm
HI}^{3/2}\over|V'_{\rm HI}|}\right\vert_{\sigma=\sigma_*}$.

\item The spectral index $n_{\rm s}=1-6\epsilon_*\ +\
2\eta_*$ and its running $\alpha_{\rm s}={2\left(4\eta_*^2-(n_{\rm
s}-1)^2\right)\over3}-2\xi_*$,
where $\epsilon\simeq{m^2_{\rm P}\over2}\left(\frac{V'_{\rm
HI}}{V_{\rm HI}}\right)^2,~\eta\simeq m^2_{\rm P}~\frac{V''_{\rm
HI}}{V_{\rm HI}}~~\mbox{and}~~\xi\simeq m_{\rm P}^4{V'_{\rm HI}
V'''_{\rm HI}\over V^2_{\rm HI}}$ and the subscript $*$ means that
the quantities are evaluated for $\sigma=\sigma_*$
\end{itemize}

If FHI is to produce the total amount of e-foldings, $N_{\rm
tot}$, needed for the resolution of the horizon and flatness
problems of standard cosmology, i.e., $N_{\rm tot}=N_{\rm
HI*}\simeq50$, we get $n_{\rm s}\sim 0.98-1$ which is just
marginally consisted with the fitting of the WMAP5 data
\cite{wmap3} by the standard power-law cosmological model
$\Lambda$CDM, according to which
\begin{equation}\label{nswmap}
n_{\rm s}=0.963^{+0.016}_{-0.015}~\Rightarrow~0.931\lesssim n_{\rm
s} \lesssim 0.991
\end{equation}
at 95$\%$ confidence level with negligible $a_{\rm s}$. However,
for $\kappa\simeq (0.01 - 0.1)$ and $N_{\rm HI*}\sim (15-20)$ we
can obtain $n_{\rm s}\simeq 0.96$. $N_{\rm tot}-N_{\rm HI*}$ can
be produced during another stage of (complementary) inflation,
realized at a lower scale. In this talk, which is based on
Ref.~\cite{mhi}, we show that MI can successfully play this role.

\section{The basics of Modular Inflation}

After the gravity mediated soft SUSY breaking, the potential which
can support MI has the form \cite{modular} $V_{\rm MI}=V_{\rm
MI0}-m_s^2s^2/2+\cdots$, with $V_{\rm MI0}=v_s(m_{3/2}m_{\rm
P})^2$ and $m_s\sim m_{3/2}$ where $m_{3/2}\sim 1~{\rm TeV}$ is
the gravitino mass, the coefficient $v_s$ is of order unity and
the ellipsis denotes terms which are expected to stabilize $V_{\rm
MI}$ at \mbox{$s\sim m_{\rm P}$} with $s$ being the canonically
normalized string modulus. In this model, inflation can be of the
slow or fast-roll type \cite{fastroll} depending on whether
$|\eta_s|=m^2_{\rm P}{|d^2V_{\rm MI}/ds^2|/V_{\rm MI}}={m_s^2/
3H_s^2}$ is lower or higher than unity, respectively. In both
cases the solution of the equation of motion of $s$ during MI is
\beq s=s_{\rm Mi}e^{F_s \Delta N_{\rm MI}}~~{\rm with}~~
F_s\equiv\sqrt{{9/4}+\left({m_s/H_s}\right)^2}-{3/2}, \label{Fs}
\eeq
with $H_s\simeq\sqrt{V_{\rm MI0}}/ \sqrt{3}m_{\rm P}$, $s_{\rm
Mi}$ the value of $s$ at the onset of MI and $\Delta N_{\rm MI}$
the number of the e-foldings obtained from $s= s_{\rm Mi}$ until a
given $s$. Through the use of Eq.~(\ref{Fs}) and considering that
the final value of $s$, $s_{\rm f}$, is close to its VEV, $s_{\rm
f}\sim m_{\rm P}$, we can estimate the total number of e-foldings
during MI, which is $ N_{\rm MI}\simeq\frac{1}{F_s}
\ln\left(\frac{m_{\rm P}}{s_{\rm Mi}}\right).$ We observe that MI
can not play successfully the role of complementary inflation for
$s_{\rm Mi}/m_{\rm P}\gtrsim0.1$.

\section{Observational Constraints}

Our double inflationary model needs to satisfy a number of
constraints which arise from:
\begin{enumerate}
\item \emph{The normalization of $P_{\cal R*}$:} We require
\cite{wmap3} $P^{1/2}_{\cal R*}\simeq\: 4.86\times 10^{-5}$.
\item \emph{The resolution of the horizon and flatness problems}:
We entail $N_{\rm HI*}+N_{\rm MI}\simeq22.6+{2\over
3}\ln{V^{1/4}_{\rm HI0}\over{1~{\rm GeV}}}+ {1\over3}\ln {T_{\rm
Mrh}\over{1~{\rm GeV}}}$ where we assumed that there is matter
domination in the inter-inflationary era ($T_{\rm Mrh}$ is the
reheat temperature after MI).
\item \emph{The Low Enough Value of $\alpha_{\rm s}$.}
Consistently with the power-law $\Lambda$CDM model we demand:
$\vert\alpha_{\rm s}\vert \ll 0.01.$
\item \emph{The naturalness of MI.} For natural MI we need:
$0.5\leq v_s\leq 10~\Rightarrow~2.45\geq {m_s\over H_s}\geq 0.55$.
\item \emph{The Nucleosynthesis Constraint.} This constraint
dictates $T_{\rm Mrh}>1\>{\rm MeV}$. In the absence of other
specified interactions, $s$ has just gravitational interactions.
Therefore, $\Gamma_s\sim m^2_s/m_{\rm P}^3$ and since $T_{\rm
Mrh}\sim \sqrt{\Gamma_s m_{\rm P}}$, we need \cite{mhi}
$m_{s}\simeq m_{3/2}\geq100\>{\rm TeV}$.

\item  \emph{The evolution of the cosmological scales.} We have to
ensure that the cosmological scales leave the horizon during FHI
and do not re-enter the horizon before the onset of MI. This can
be achieved \cite{anupam} if $N_{\rm HI*}\gtrsim N^{\rm min}_{\rm
HI*}\simeq3.9+{1\over 6}\ln {V_{\rm HI0}\over V_{\rm MI0}}\sim
10.$

\item \emph{The evolution of $s$ before the onset of MI.} (i) If
$m_s|_{\rm eff}=0$, we assume that FHI lasts long enough so that
$s$ is completely randomized. We further require that all $s$'s
belong to the randomization region \cite{chun} with equal
possibility, i.e., $V_{\rm MI0}\lesssim H_{\rm HI0}^4$ where
$H_{\rm HI0}=\sqrt{V_{\rm HI0}}/ \sqrt{3}m_{\rm P}$. (ii) If
$m_s|_{\rm eff}\neq0$, we assume that $s$ is decoupled from the
visible sector superfields both in K\" ahler potential and
superpotential and has canonical K\"{a}hler potential,
$K_s=s^2/2$. Therefore the value $s_{\rm min}$ at which the SUGRA
potential has a minimum is $s_{\rm min}=0$. We obtain for the
value of $s$ at the onset of MI: $s_{\rm Mi}\simeq m_{\rm
P}\left({V_{\rm MI0 }/V_{\rm HI0 }}\right)^{1/4}e^{-3N_{\rm
HI}/2}$ where $s_{\rm Hi}\simeq m_{\rm P}$ is the value of $s$ at
the onset of FHI and $N_{\rm HI}$ the total number of e-foldings
during FHI.

\item \emph{The homogeneity of the present universe.} If
$\left.\delta s\right|_{\rm MI}$ [$\left.\delta s\right|_{\rm
HMI}$] are the quantum fluctuations of $s$ during MI [FHI which
enter the horizon of MI], we require $ s_{\rm Mi}> \left.\delta
s\right|_{\rm MI}\simeq H_s/2\pi$ and $s_{\rm Mi}> \left.\delta
s\right|_{\rm HMI}$. (i) If $m_s|_{\rm eff}=0$, $\left.\delta
s\right|_{\rm HMI}\simeq H_{\rm HI0}/2\pi\gg\left.\delta
s\right|_{\rm MI}$. (ii) If $m_s|_{\rm eff}\neq0$, $\left.\delta
s\right|_{\rm HMI}\simeq H_s/3^{1/4}2\pi<\left.\delta
s\right|_{\rm MI}$ and so, $s_{\rm Mi}>\left.\delta s\right|_{\rm
MI}\simeq H_s/2\pi~\Rightarrow~N_{\rm HI}\leq N_{\rm HI}^{\rm
max}$ where $N_{\rm HI}^{\rm max}=-{2\over3}\ln {\left(V_{\rm
HI0}V_{\rm MI0}\right)^{1/4}\over 2\sqrt{3}\pi m_{\rm
P}^2}\sim(15-18)$. This result signalizes an ugly tuning since it
would be more reasonable FHI has a long duration due to the
flatness of $V_{\rm HI}$. This could be evaded if we had $s_{\rm
min}\neq0$ (as in Ref.~\cite{referee}).
\end{enumerate}

\section{Results and Conclusions}

In our numerical investigation, we take  ${\sf N}=2$ and
$m_{3/2}=m_s=100~{\rm TeV}$ which results to $T_{\rm Mrh}=1.5~{\rm
MeV}$. Our results are displayed in Table 1 for $n_{\rm s}=0.963$,
$m_s|_{\rm eff}=0$ or $m_s|_{\rm eff}\neq0$ and selected
$\kappa$'s which delineate the allowed regions. For $m_s|_{\rm
eff}=0$ we place $s_{\rm Mi}/m_{\rm P}=0.01$. This choice
signalizes a very mild tuning (see point 7). For $m_s|_{\rm
eff}\neq0$, $s_{\rm Mi}$ is evaluated dynamically (see point 7).
However, due to our ignorance of $N_{\rm HI}$, we can derive a
maximal [minimal] $m_s/H_s$ which corresponds to $N_{\rm
HI}=N^{\rm max}_{\rm HI}$ [$N_{\rm HI}=N_{\rm HI*}$].

\begin{table}[t]
\centering
\begin{tabular}{lclllccllll} \hline
&& \tablehead{3}{c}{b}{$m_s|_{\rm eff}=0$}&&&
\tablehead{4}{c}{b}{$m_s|_{\rm eff}\neq0$}
\\ \hline\hline
$\kappa$ && $0.04$ &$0.09$&$0.14$&&&$0.0028$
&$0.006$&$0.085$&$0.14$\\\hline
$M/10^{16}~{\rm GeV}$ && $0.87$&$0.98$&$1.07$&&&
$0.74$&$0.8$&$0.97$&$1.07$\\
$\sigma_*/10^{16}~{\rm GeV}$ &&$12.1$ &$20.93$&$25.88$&&&$1.56$
&$2.26$&$20.1$&$25.88$\\ \hline
$N_{\rm HI*}$ &&$22.6$  &$16.12$&$11.9$&&&$8.4$ &$17.4$&$16.5$&$11.9$\\
$-\alpha_{\rm s}/10^{-3}$ && $2$ &$5$&$10$ &&& $2.4$&$1.5$&$4.8$&$10$\\
\hline
$N_{\rm MI}$ &&$21.2$& $28$&$32.5$&&&$34.1$ &$25.7$&$27.6$&$32.5$\\
$m_s/H_s$ &&$0.8$& $0.72$&$0.67$&&&$1.44-1.96$& $2.35$&$2.25$&$1.78-2.02$\\
\hline
\end{tabular}
\caption{\sl Input and output parameters of our scenario which are
consistent with the requirements 1-8 for $n_{\rm s}=0.963$ and
selected $\kappa$'s, when the inflaton of MI does [does not]
acquire effective mass ($m_s|_{\rm eff}\neq0$ [$m_s|_{\rm
eff}=0$]).}\label{tab}
\end{table}

We observe that (i) for $m_s|_{\rm eff}=0$ [$m_s|_{\rm
eff}\neq0$], the lowest $\kappa$'s are derived from the condition
7 [6] and therefore, lower $\kappa$'s are allowed for $m_s|_{\rm
eff}\neq0$; (ii) the upper $\kappa$'s come from the condition 3;
(iii) for $m_s|_{\rm eff}=0$ [$m_s|_{\rm eff}\neq0$], MI is of
slow [fast]-roll type since $m_s/H_s\sim (0.6-0.8)$ [$m_s/H_s\sim
(1.4-2.35)$]; (v) for $m_s|_{\rm eff}\neq0$ FHI is constrained to
be of short duration since $N_{\rm HI}\leq N^{\rm max}_{\rm
HI}\simeq(16-17)$ and as a consequence, the region
$0.006\lesssim\kappa\lesssim0.085$ is disallowed; (vi) in both
cases, the allowed $M$'s increase with $\kappa$'s but remain
slightly below the GUT scale, $M_{\rm
GUT}\simeq2.86\cdot10^{16}~{\rm GeV}$. In total, we obtain
$0.04\lesssim\kappa\lesssim0.14$
[$0.0028\lesssim\kappa\lesssim0.006$ and
$0.085\lesssim\kappa\lesssim0.14$] for $m_s|_{\rm eff}=0$
[$m_s|_{\rm eff}\neq0$].

In conclusion, we showed that the results on $n_{\rm s}$ within
FHI can be reconciled with data if FHI is followed by MI realized
by a string modulus $s$. Acceptable $n_{\rm s}$'s can be obtained
by restricting $N_{\rm HI*}$. The most natural version of this
scenario is realized when $s$ remains massless before MI.

\begin{theacknowledgments}
This work was supported by the Spanish MEC project FPA2006-13825
and the project P07FQM02962 funded by the ``Junta de Andalucia''
and the FP6 Marie Curie Excellence Grant MEXT-CT-2004-014297.
\end{theacknowledgments}

\newcommand{\arxiv}[1]{{\tt arXiv:#1}}

\newcommand{\hepph}[1]{{\tt hep-ph/#1}}
\newcommand{\hepex}[1]{{\tt hep-ex/#1}}
\newcommand{\astroph}[1]{{\tt astro-ph/#1}}
\newcommand{\hepth}[1]{{\tt hep-th/#1}}
\newcommand{\grqc}[1]{{\tt gr-qc/#1}}
\newcommand{\etal}{{\it et al.\/}}

\newcommand\astp[3]{\emph{ Astropart.\ Phys.\ }{\bf #1}, #3 (#2)}

\newcommand\apj[3]
        {\emph{ Astropart.\ J. \ }{\bf #1}, #3 (#2)}

\newcommand\jhep[3]
        {\emph{ J. High Energy Phys.\ }{\bf #1}, #3 (#2)}

\newcommand\jcap[3]
        {\emph{ J. Cosmol. Astropart. Phys.\ }{\bf #1}, #3 (#2)}

\newcommand\npb[3]
        {\emph{ Nucl.\ Phys.\ }{\bf B#1}, #3 (#2)}

\newcommand\plb[3]
        {\emph{Phys.\ Lett.\ }{B \bf #1}, #3 (#2)}

\newcommand\prd[3]
        {\emph{ Phys.\ Rev.\ }{D \bf #1}, #3 (#2)}

\newcommand\prep[3]
        {\emph{ Phys.\ Rep.\ }{\bf #1}, #3 (#2)}
\newcommand\prl[3]
        {\emph{ Phys.\ Rev.\ Lett.\ }{\bf #1}, #3 (#2)}


\end{document}